\documentclass[12pt]{article}

\usepackage{times}
\usepackage{graphicx}
\usepackage{gensymb}
\usepackage{setspace} 

\usepackage{color}

\usepackage{booktabs}

\usepackage{lineno}

\newcounter{lastnote}

\title{Moral attitudes and willingness to induce cognitive enhancement and repair with brain stimulation}

\author{John D. Medaglia$^{1,2\ast}$, David Yaden$^{3}$, Chelsea Helion$^{4}$, Madeline Haslam$^{5}$\\
	\normalsize{$^{1}$Department of Psychology, Drexel University}\\
	\normalsize{Philadelphia, PA, 19104, USA}\\
	\normalsize{$^{2}$Department of Neurology, Perelman School of Medicine, University of Pennsylvania}\\
	\normalsize{Philadelphia, PA, 19104, USA}\\
	\normalsize{$^{3}$Department of Psychology, University of Pennsylvania}\\
	\normalsize{Philadelphia, PA, 19104, USA}\\	
	\normalsize{$^{4}$Department of Psychology, Columbia University}\\
	\normalsize{Philadelphia, PA, 19104, USA}\\	
	\normalsize{$^{5}$Department of Psychology, Washington College}\\
	\normalsize{Chestertown, MD, 21620, USA}\\	
\normalsize{$\ast$ To whom correspondence should be addressed: john.d.medaglia@drexel.edu}\\
	}

\begin{document} 


\baselineskip24pt


\maketitle 

\clearpage
\newpage
\singlespacing


\section*{Abstract}
The availability of technological means to enhance and repair human cognitive function raises questions about the perceived morality of their use. In this study, we administered a survey to the public in which subjects were asked to report how willing they would be to enhance and/or repair specific cognitive abilities. Among 894 responders, we found that subjects were more willing to use technologies to repair other people than themselves, and especially to enhance or repair functions more ``core" to authentic identity in others. Subjects' ratings of the moral acceptability of specific uses was related to their reported willingness to use brain stimulation. These findings suggest that the public endorses an altruistic approach to applying brain stimulation for cognitive gains. Further, this study establishes a basis to guide moral psychological studies of cognitive modification and social processes that guide attitudes toward and uses of brain stimulation.
\clearpage
\newpage

\section*{Author Contributions}
JDM conceptualized the study, analyzed the data, and wrote the manuscript. DY helped to conceptualize the study and contributed to manuscript writing. CH helped to conceptualize the study, edit manuscript drafts, and interpret findings. MH contributed to the study design, item creation, and survey creation and administration.
\clearpage
\newpage

\section*{Introduction}
Humans enhance cognitive functions through physical and cognitive training \cite{ngandu20152}, pharmaceuticals \cite{franke2014substances}, and invasive and noninvasive neuromodulation \cite{snowball2013long}. With increasing public availability of potentially effective cognitive modification tools has emerged a ``do-it-yourself" (DIY) noninvasive brain stimulation movement \cite{wexler2016practices}. Within the DIY movement, people use technologies such as transcranial direct current stimulation in attempts to treat depression and other psychological problems or to enhance cognition \cite{wexler2017who}. Beyond this movement, we can reasonably anticipate that with the combination of easy access to home manufacturing technologies \cite{gibson2014additive} and open-source information that laypersons may build ever-more effective tools for neuromodulation at home. In addition, technologies such as deep-brain stimulation provide powerful invasive means to influence human cognition \cite{okun2009cognition}. In this context, researchers and ethicists have debated how we should study public use of brain stimulation \cite{jwa2015early,wexler2016practices,wexler2017crowdsourced}, communicate information with consumers \cite{wurzman2016open}, prioritize who receives interventions \cite{hamilton2011rethinking}, and regulate available technologies \cite{fitz2013challenge,wexler2016practices}. 

Experiments suggest that the public has a sophisticated perspectives and sensitivity to major debates in cognitive enhancement research \cite{fitz2014public}. However, while the (DIY) culture is rapidly growing with unique opportunities for scientific discovery \cite{jwa2015early,wexler2016practices,wexler2017crowdsourced}, and more invasive techniques are being applied for a variety of disorders, we know relatively little about public perspectives and moral appraisals about human enhancement techniques more broadly. Several major variables could affect human willingness to use certain technologies to influence cognition. Here, we focus on whether people are willing to use a hypothetical brain stimulation technology to influence cognitive features that are more or less core to the ``authentic self" \cite{harter2002authenticity}. We additionally examined whether subjects are more willing to repair or enhance features, and more willing to influence others \emph{versus} themselves. Finally, we considered the extent to which judgments of moral acceptability were associated with the willingness to apply certain uses in general. Examining these variables could begin to identify what guides intuitive judgments about brain stimulation for cognition. It could further establish the contributions of moral evaluations to public attitudes about brain stimulation uses.

Some previous research points to the potential relevance of these variables. Research investigating consumers' willingness to take pharmaceuticals indicated that young, healthy adults were reluctant to enhance traits believed to be more fundamental to one's self-identity (the ``authentic self") than others \cite{riis2008preferences,cabrera2015reasons}. However, it is unknown whether similar preferences generalize to non-pharmaceutical interventions such as brain stimulation. In addition, individuals may view interventions used to repair (i.e., therapeutically) rather than enhance performance more acceptable than others \cite{cabrera2015empirical} when moving individuals toward normative functioning. Further, individuals may sometimes seek to gain performance advantages for themselves rather than for others in well-controlled experimental contexts \cite{batson1999moral}. However, the extent to which this applies to potentially advantageous or altruistic applications of brain stimulation has not been examined. Moreover, moral acceptability -- whether engaging in an action is thought to be the right thing to do -- was strongly related to the desire to legalize enhancement as opposed to personally taking the enhancement \cite{riis2008preferences}. This suggests that individuals may engage in moral reasoning when considering institutional action about enhancements, but not on limited uses on an individual basis. However, whether moral evaluations contribute to willingness to apply cognitive enhancement and repair more generally has not been thoroughly evaluated.

In a within-between subjects design, we examined individuals' willingness to apply a limited supply of a hypothetical brain stimulation device called ``Ceremode'' to enhance or repair cognitive features in themselves or others. We hypothesized that subjects would be less willing to enhance or repair traits they perceive as more fundamental to ``authentic" self-identity \cite{riis2008preferences}. We further predicted that in the zero-sum vignettes, subjects would prioritize enhancing or repairing cognitive functions in themselves rather than others. We anticipated that subjects would be more willing to repair than enhance cognitive functions, and similarly willing to enhance traits less fundamental (``peripheral") to authentic identity in themselves and others, but more willing to enhance more fundamental (``core") traits in others. We expected moral acceptability to be positively related to subjects’ likelihood ratings across conditions. Finally, we anticipated that demographic variables associated with particular ideological positions could be associated with willingness to apply certain uses. Specifically, we speculated that politically more conservative individuals would be less willing to enhance or repair individuals, and that this effect would be partially mediated by Openness to Experience \cite{gerber2011big}.

\section*{Open Science Framework Pre-registration}
	We are sensitive to increasing concerns about scientific transparency, academic honesty, the influence of ``perverse" incentives on researchers \cite{stephan2012research,edwards2017academic}, and the complexity of interactions between variables that influence psychological research \cite{maxwell2015psychology}. We considered it important to constrain the focus of our work to test hypotheses that were discussed among the coauthors ahead of time and only to test the predictions that we made before looking at any data. Thus, prior to any data analysis, we pre-registered the current confirmatory hypotheses for this study \emph{a priori} in the Open Science Framework (www.osf.io). Thus, the primary results in the current manuscript presents only the procedures and results proposed in our initially intended research as protection against implicit or explicit researcher biases during data analysis. 
	
	\subsection*{Subjects}
	In two separate survey administrations (one each for the  ``self" and ``other" conditions, between subjects), 500 subjects for each condition were solicited via Amazon Mechanical Turk (M-Turk). M-Turk allows researchers to collect high-quality data rapidly among demographically diverse individuals \cite{buhrmester2011amazon}. The participants were required to have completed at least 100 tasks previously and maintain at least a 95\% approval rating on M-Turk. The study was approved by the Institutional Review Board at the University of Pennsylvania and all participants provided informed consent prior to participating in the study.
	
	\subsection*{Survey design}
	We administered two surveys (``self" and ``other"): one to assess using technology on one's self (``self survey") and another to assess using technologies on others (``other survey"). In these surveys, we assessed the associations of three primary variables with subjects' likelihood of using a fictitious technology called \emph{``Ceremode''} (see description below) to influence specific cognitive functions. Within each survey, we asked subjects to consider whether they would enhance or repair, specific cognitive functions that were selected to be either more core or peripheral to authentic self-identity based on a previous study \cite{riis2008preferences}. Then, we asked subjects to consider how they would distribute \emph{Ceremode} to affect specific cognitive functions in a zero-sum scenario. Next, we asked subjects to rate the moral acceptability of specific uses of \emph{Ceremode}. Afterwards, we asked subjects to respond to items measuring demographic variables, personality, and political ideologies. Thus, the self/other condition was between subjects, whereas the enhance/repair and core/peripheral factors were within subjects. Specific procedures for each section are described below.
	
		\paragraph{Self Survey design: giving \emph{Ceremode} to the self}
		This survey began with definitions of each of 16 terms referring to specific cognitive functions from a prior study that established them as more core or peripheral to human self-identity \cite{riis2008preferences}. Specifically, we selected the eight most ``core" functions (Kindness, Empathy, Self-confidence, Mood, Motivation, Social Comfort, Relaxation,	and Emotional Recovery) and the eight most ``peripheral" functions (Reflexes, Rote Memory, Wakefulness, Foreign language ability, Math ability, Episodic memory, Concentration,
		and Music ability).
		
		Subjects were instructed to respond as follows:
		
		\begin{quote}``In a breakthrough discovery, scientists have created a brain stimulation device called Ceremode that can be used on individuals to change cognitive abilities. There are several different versions of Ceremode. Each version can affect only one type of cognitive ability at a time without side effects. In addition, each version of Ceremode can be used in either healthy individuals or individuals with brain injuries to change cognitive abilities.
		
		\textbf{Imagine that you are responsible for deciding whether to use Ceremode on yourself to affect your cognitive functions. With this in mind, your task is to read sentences describing the effects of each version of Ceremode. Then, decide how likely you would be to give that version to yourself.} Some versions of Ceremode will be used to enhance cognitive functions while you are healthy. Other versions of Ceremode will be used to repair cognitive functions if you suffer a brain injury. You will rate each item from 1 (Extremely Unlikely) to 5 (Extremely Likely). Importantly, you cannot receive every version of Ceremode because only so many versions of each type of Ceremode exist. \textbf{Considering that Ceremode comes in a limited supply}, please use the entire range from 1-5 when making your decisions."\end{quote}

		Then, we presented 32 items. Half of the items asked whether subjects would be likely to enhance their functions while they are healthy, and half of the items asked whether subjects would be likely to repair functions if they had brain injuries:
		
		\begin{quote}
		``This version of Ceremode can be used on you \emph{$<$while you are healthy/if you have brain injuries$>$} to \emph{$<$enhance/repair$>$} cognitive abilities. If you are stimulated by this Ceremode you will receive improved \emph{$<$cognitive function$>$ with no side effects}.
		
		I would be $_{-----}$ to give this version of Ceremode."
		\end{quote}
		
		If the prompt included ``while you are healthy'', then it was always followed by ``enhance''. If the prompt included ``if you have brain injuries'', then it was always followed by ``repair''.
		
		We included two attention checks in this portion of the survey. Each attention check stated:
		\begin{quote}
			``This is an attention check. If you are paying attention please select Extremely Likely."
		\end{quote}
		
		For each item, subjects responded on a Likert scale from among the responses (1) Extremely Unlikely; (2) Somewhat Unlikely; (3) Neutral; (4) Somewhat Likely; and (5) Extremely Likely. The item order was presented uniformly at random for each participant.
		
		\paragraph{Distributing Ceremode to the self with a zero-sum constraint}
		After assessing subjects' estimated likelihood of administering \emph{Ceremode} to enhance or repair specific functions, we asked individuals to distribute a limited number of \emph{Ceremode} versions to enhance or repair various cognitive functions (within subjects). That is, subjects were shown a list of various cognitive functions and asked to distribute a limited number of units to the functions that they would most want enhanced or repaired. 
		
		In the ``enhance" condition, subjects were asked:
		\begin{quote}
			``Now imagine that you receive a special version of Ceremode. This version can enhance your functions right now. It can enhance many functions. However, it has a limited amount of energy, which affects how much you can enhance each cognitive function. For example, you could set it to enhance all functions a little bit, or a single function completely, or anywhere in between.
			
			Please read the whole list carefully before making your choices and please use 16 total units of energy (that is, the total of energy across all items should equal 16)."
		\end{quote}

		In the ``repair" condition, subjects were asked:
		\begin{quote}
			``Now imagine that you receive a different special version of Ceremode. This version can repair lost functions if you suffer a brain injury. It can repair many functions. However, it also has a limited amount of energy, which affects how much you can repair each cognitive function. For example, you could set it to repair all functions a little bit, or a single function completely, or anywhere in between.
			
			Please read the whole list carefully before making your choices and please use 16 total units of energy (that is, the total of energy across all items should equal 16)."
		\end{quote}
		
		Underneath the prompt for each condition, the subjects typed the number of units they would allocate to each of the 16 cognitive functions into a box next to the name of each cognitive function. Responses were fixed so that subjects had to allocate exactly 16 total units (i.e., the sum of allocated units across all functions equaled 16) before proceeding. 
		
		\paragraph{Moral acceptability of using \emph{Ceremode} on the self}
		After assessing subjects' distribution of \emph{Ceremode} to various cognitive functions, we asked the extent to which they thought specific uses of this hypothetical technology on the self were morally acceptable. 
		
		For the \emph{enhance} condition, we asked subjects:
		\begin{quote}
			``Now consider that there is no limit to how many versions of Ceremode you can use to enhance your cognitive abilities. How morally acceptable are each of the following uses of Ceremode to enhance your cognitive abilities right now?"
		\end{quote}
	
		For the \emph{repair} condition, we asked subjects:
		\begin{quote}
			``Now consider that there is no limit to how many versions of Ceremode you can use to repair your cognitive abilities. How morally acceptable are each of the following uses of Ceremode to repair your abilities if you suffer a brain injury?"
		\end{quote}
			
				Then, for each of the two conditions, for each of the sixteen cognitive functions, subjects were prompted with a series of items in a randomized order with the format:
				
		\begin{quote}
			``It is morally acceptable to $<$enhance/repair$>$ your $<$cognitive function$>$."
		\end{quote}
		
		For each item, subjects responded on a Likert scale from among the responses (1) Strongly disagree; (2) Somewhat disagree; (3) Neutral; (4) Somewhat agree, and (5) Strongly agree. The item order was presented uniformly at random for each participant.

		\paragraph{Other Survey design: giving \emph{Ceremode} to others}
		The survey measuring attitudes toward other-enhancement was structured identically to the self survey, except that the instructions and prompts were changed to ask a different sampel of individuals to think about using Ceremode on others instead of themselves.

		\paragraph{Moral acceptability of using \emph{Ceremode} on others}\
		After assessing subjects' distribution of \emph{Ceremode}, we asked the extent to which they thought specific uses of this hypothetical technology on others were morally acceptable. 
		
		For the \emph{enhance} condition, we asked subjects:
		\begin{quote}
			``Now consider that there is no limit to how many versions of Ceremode you can use to enhance cognitive abilities. How morally acceptable are each of the following uses of Ceremode to enhance cognitive abilities in healthy individuals?"
		\end{quote}	
		
		For the \emph{repair} condition, we asked subjects:
		\begin{quote}
			``Now consider that there is no limit to how many versions of Ceremode you can use to repair cognitive abilities. How morally acceptable are each of the following uses of Ceremode to repair abilities after individuals have suffered brain injuries?"
		\end{quote}
		
		Then, for each of the two conditions, for each of the sixteen cognitive functions, subjects were prompted with a series of items in a randomized order with the format:
		
		\begin{quote}
			``It is morally acceptable to $<$enhance/repair$>$ your $<$cognitive function$>$."
		\end{quote}
		
		For each item, subjects responded on a Likert scale from among the responses (1) Strongly disagree; (2) Somewhat disagree; (3) Neutral; (4) Somewhat agree, and (5) Strongly agree. The item order was presented uniformly at random for each participant. 
		
		\paragraph{Big Five Inventory-10 (BFI-10)}
		Following the Moral Acceptability section in each of the \emph{enhance} and \emph{repair} conditions, subjects completed the BFI-10 to provide validated and time-efficient measurements of the Big Five personality traits \cite{rammstedt2007measuring}.
		
		\paragraph{Neoliberal Beliefs Inventory (NBI)}
		Subjects then completed the NBI \cite{bay2015tracking}, which is a reliable and valid 25-item measure of neoliberal beliefs. The NBI contains four factors that entail beliefs about system inequality, competition, personal wherewithal, and government interference.
		
		\paragraph{Demographics}
		Subjects reported their gender, age, race/ethnicity, education, income, field of study, employment status, and political party. In addition, subjects reported their religious upbringing, current religious tradition, degree of religiosity, degree of spirituality, church or other religious meeting attendance, time spent in religious/spiritual activities, relationship status,  and an reporting whether they thought human minds are more like computer or souls.

\subsection*{Statistical analyses} 
Analyses were conducted using multilevel modeling with maximum-likelihood estimation \cite{baayen2008mixed} implemented in the lme4 v.1.1-9  \cite{bates2014lme4} package of R version 3.2.1 \cite{r2016software}. This technique allows for a classical regression analysis to be performed on repeated measures data by accounting for the non-independence of observations collected from each participant in a within-subjects design, without resorting to computing separate regression equations for each subject \cite{lorch1990regression,baayen2008mixed,baayen2008analyzing}. Multilevel modeling also accounts for violations of the sphericity assumption by modeling heteroscedasticity in the data when necessary, improving statistical power over other methods commonly employed for analyzing repeated-measures data. Specifically, we examined the Self/Other, Enhance/Repair, and Core/Peripheral variables as discrete independent variables, age, education, and gender as covariates, and included random effects for subjects. We examined political affiliation and moral acceptability as independent variables in separate main effects models to isolate these effects.

\section*{Results}
\paragraph*{Final sample}
A total of 1000 subjects enrolled to take the surveys, half for the ``self survey" and half for the ``other survey". In the self survey, we excluded 30 subjects who did not provide complete data and an additional 21 subjects failed at least one attention check item and were excluded. In the ``other" survey, we excluded 31 subjects did not provide complete data and an additional 17 subjects failed at least one attention check item. The final sample included 894 subjects, with 449 to the ``self" survey and 445 to the ``other" survey. The total sample consisted of 443 females, a mean age of 37.4 years of age (SD = 11.55), and the modal education level was a Bachelor's degree.

\paragraph*{Main effects}
The results of the main effects model can be observed in Table~\ref{table:maineffects}. Across the sample, older individuals reported a lower willingness to use any kind of brain stimulation. Two of the three primary independent variables  (enhancement \emph{versus} repair, and self \emph{versus} other) were significant. The degree of cognitive fundamentalness (``core" \emph{versus} ``peripheral") was not.

\begin{table}[ht]
	\caption{\textbf{Main effects model}} 
	\label{table:maineffects}
	
	\begin{tabular}{l c c c c c} \toprule
		
		{} & {Estimate} & {Std. Error} & {df} & {t-value} & {\emph{p}} \\ \midrule
		
		Intercept  & 4.214 & 0.126 & 898 & 33.223 &  $<$0.001\\
		Age  	   & -1.137 & 0.002 & 889 & -5.406 &  $<$0.001\\
		Education  & -0.036 & 0.034 & 889 & -1.053 &  0.293\\
		Gender     & 0.020 & 0.048 & 889 & 0.412 &  0.680\\
		Self/Other & -0.129 & 0.048 & 889 & -2.666 &  0.008\\
		Core/Peripheral & 0.006 & 0.013 & 2771 & 0.478 &  0.633\\
		Enhance/Repair & -0.396 & 0.013 & 2771 & -30.387 &  $<$0.001\\
		\bottomrule
		
	\end{tabular}
\end{table}

\begin{figure}[h!]
	\centerline{\includegraphics[width=3.5in]{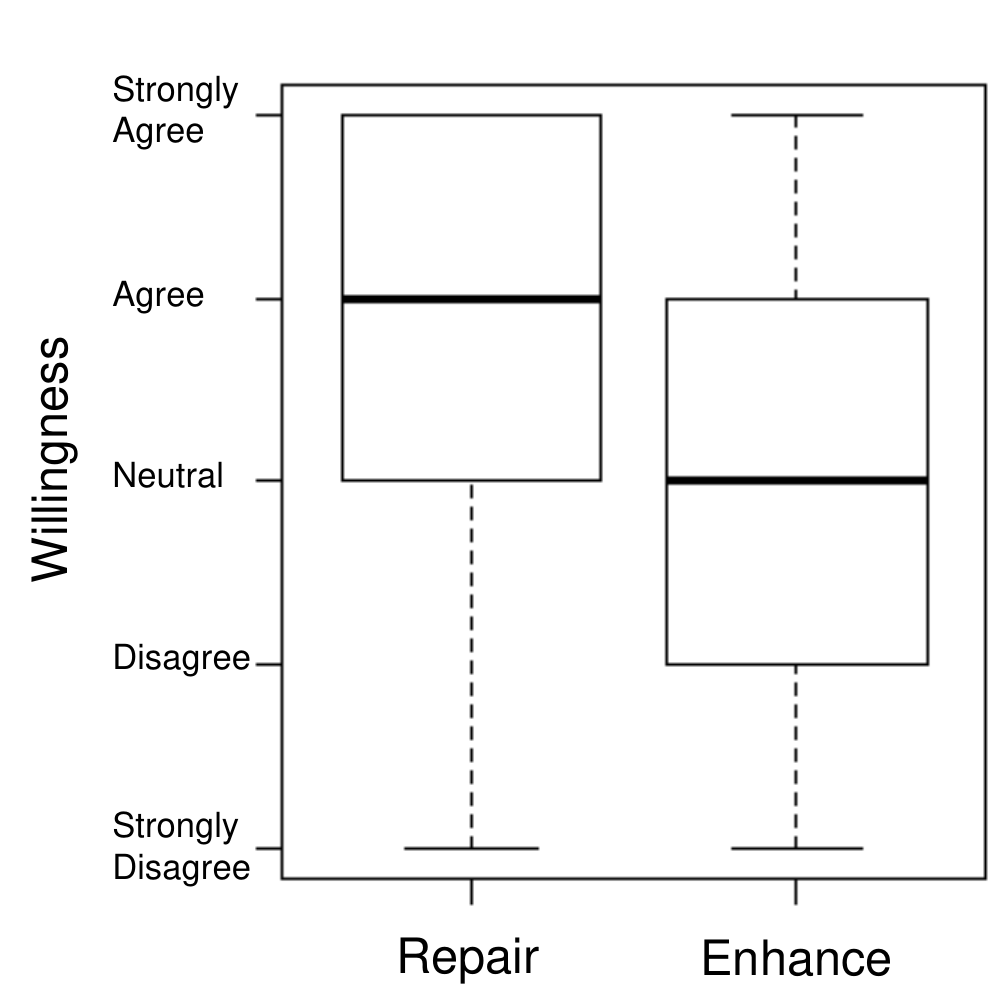}}
	\caption{\textbf{People are more willing to repair than enhance.} Overall, subjects expressed relatively
		increased willingness to use Ceremode to repair rather than enhance cognitive functions. The top and bottom horizontal hash marks represent the maximum and minimum values, respectively. The top and bottom of the box represent the third and first quartiles, respectively. The thick horizontal line represents the median value in each condition.
		Circles represent the condition means in the mixed effects model.}\label{fig:enhancerepair}
	\centering
\end{figure}
\clearpage
\newpage

\begin{figure}[h!]
	\centerline{\includegraphics[width=3.5in]{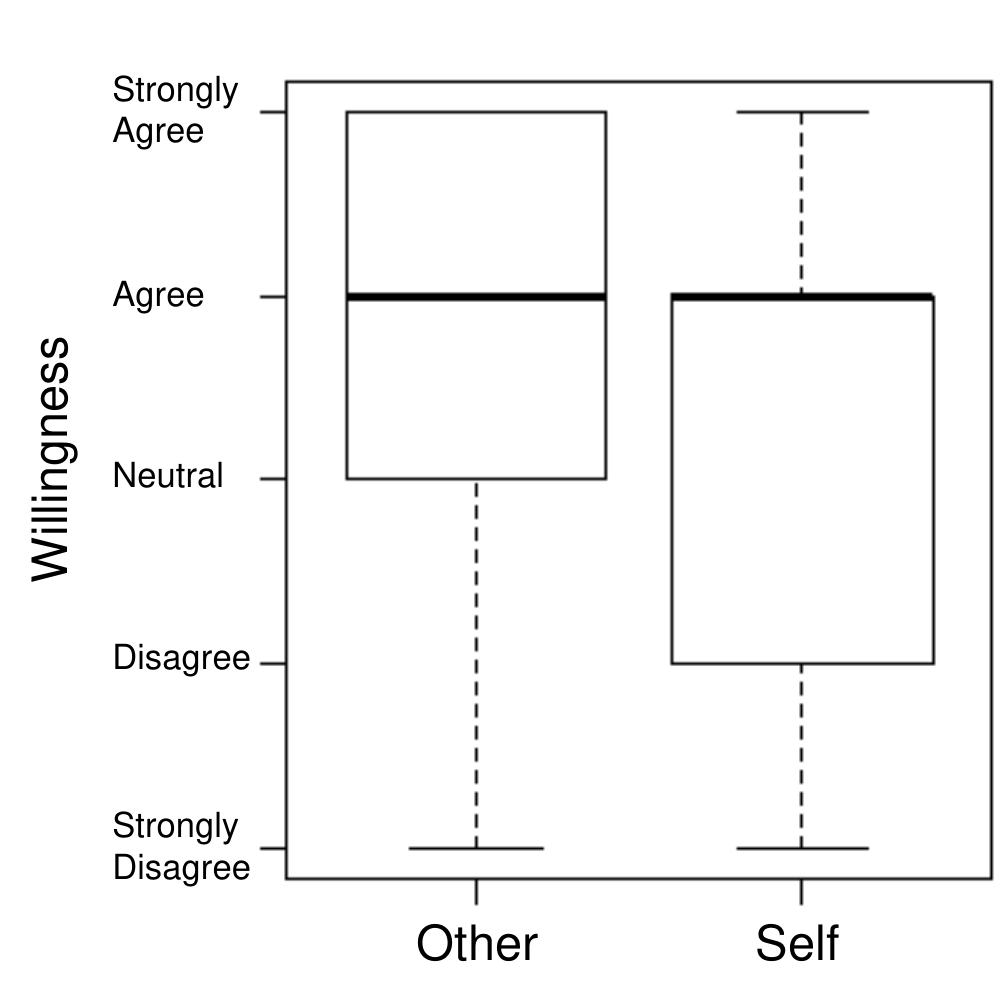}}
	\caption{\textbf{People are more willing to influence others than themselves.} Across all items,
		subjects reported a tendency to use Ceremode on others than themselves. The top and bottom horizontal hash marks represent the maximum and minimum values, respectively. The top and bottom of the box represent the third and first quartiles, respectively. The thick horizontal line represents the median value in each condition.}\label{fig:selfother}
	\centering
\end{figure}

\begin{figure}[h!]
	\centerline{\includegraphics[width=3.5in]{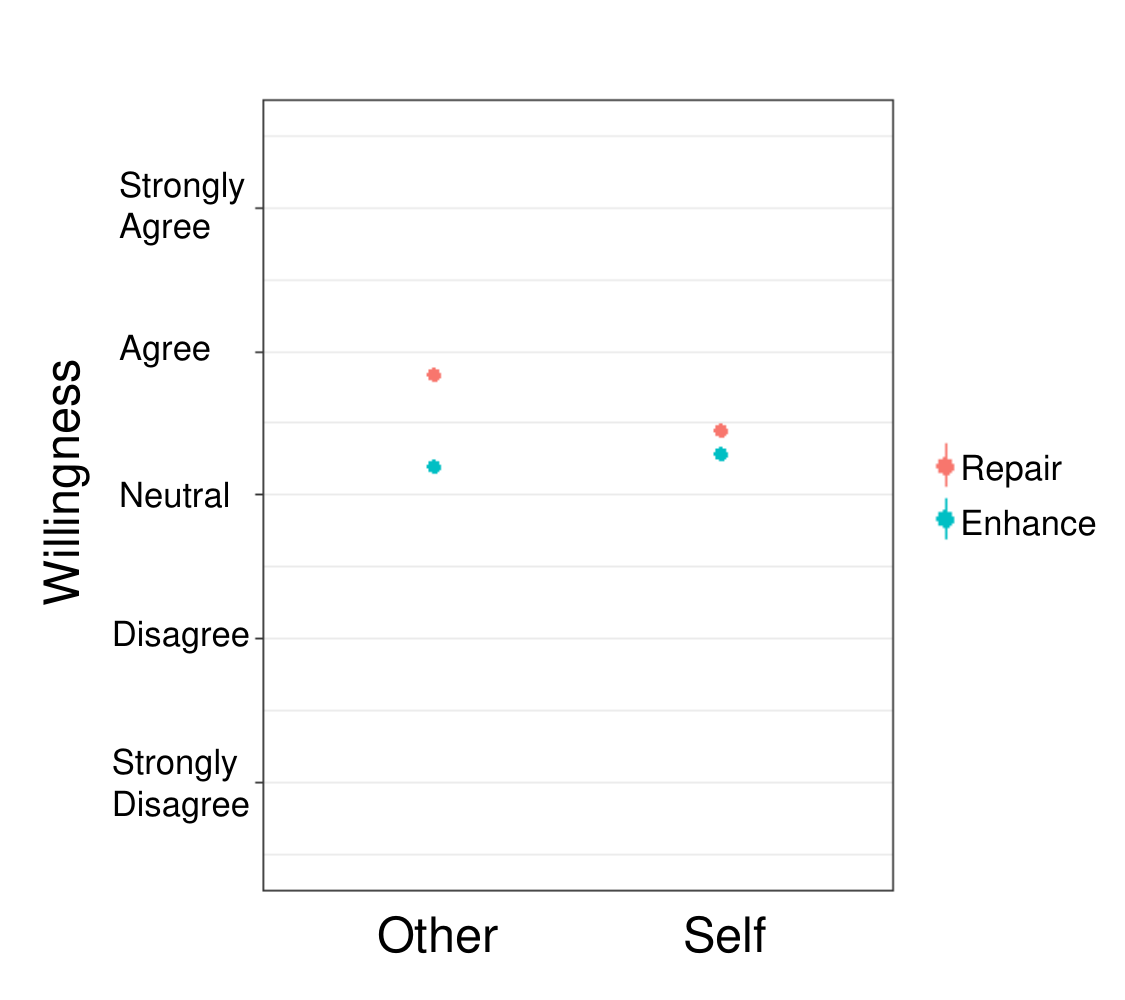}}
	\caption{\textbf{The significant interaction between the Self/Other and Enhance/Repair factors.} We observed that repair was favored over enhancement overall, and that this is especially the case when considering to use Ceremode on others. Circles represent the condition means; standard errors of the mean too small to observe at this scale.}\label{fig:selfother_enhancerepair} 
\end{figure}

\begin{figure}[h!]
	\centerline{\includegraphics[width=3.5in]{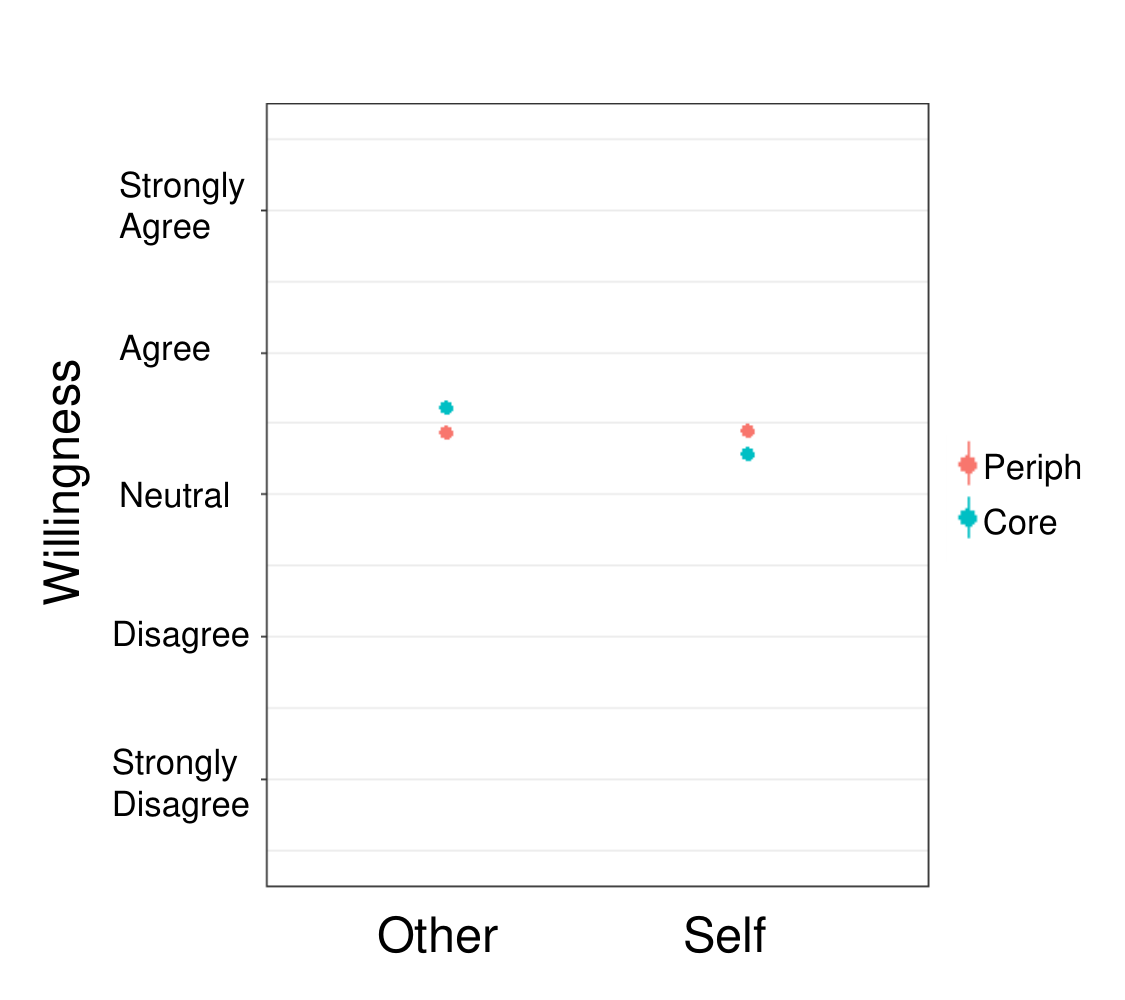}}
	\caption{\textbf{A significant interaction between Self/Other and Core/Peripheral factors.} We observed that subjects were similarly willing to use Ceremode to enhance or repair peripheral features in themselves and others. However, subjects were much more willing to use Ceremode to influence core features in others. Circles represent the condition means; standard errors of the mean too small to observe at this scale.} \label{fig:selfother_coreperiph}
	\centering
\end{figure}

The results of the mixed effects model including interactions can be viewed in Table~\ref{table:interactions}. As anticipated, the self-/other factor interacted with the cognitive fundamentalness and enhance/repair factor. Specifically, subjects were especially more likely to report willingness to use Ceremode to repair others than any other condition. In addition, they were more likely to report willingness to repair rather than enhance themselves. Among the four conditions, they were most willing to repair functions in others (See Fig.~\ref{fig:selfother_enhancerepair}), and also especially willing to enhance or repair core functions in others (See Fig.~\ref{fig:selfother_coreperiph}).

\begin{table}[ht]
	\caption{\textbf{Interaction model}} \label{table:interactions}
	
	\begin{tabular}{l c c c c c} \toprule
		
		{} & {Estimate} & {Std. Error} & {df} & {t-value} & {\emph{p}} \\ \midrule
		
		Intercept  & 4.214 & 0.126 & 908 & 33.385 &  $<$0.001\\
		Age  	   & -1.137 & 0.002 & 889 & -5.406 &  $<$0.001\\
		Education  & -0.036 & 0.034 & 889 & -1.053 &  0.293\\
		Gender     & 0.020 & 0.048 & 889 & 0.412 &  0.680\\
		Self/Other & -0.191 & 0.052 & 1160 & -3.705 &  $<$0.001\\
		Core/Peripheral & 0.179 & 0.018 & 2771 & 9.793 &  $<$0.001\\
		Enhance/Repair & -6.317 & 0.018 & 2771 & -34.521 &  $<$0.001\\
		Self/Other*Core/Peripheral & -0.344 & 0.026 & 2771 & -13.339 &  $<$0.001\\
		Self/Other*Enhance/Repair & 0.470 & 0.026 & 2771 & 18.185 &  $<$0.001\\
		\bottomrule
		
	\end{tabular}
\end{table}

\clearpage
\newpage

\begin{figure}[h!]
	\centerline{\includegraphics[width=4in]{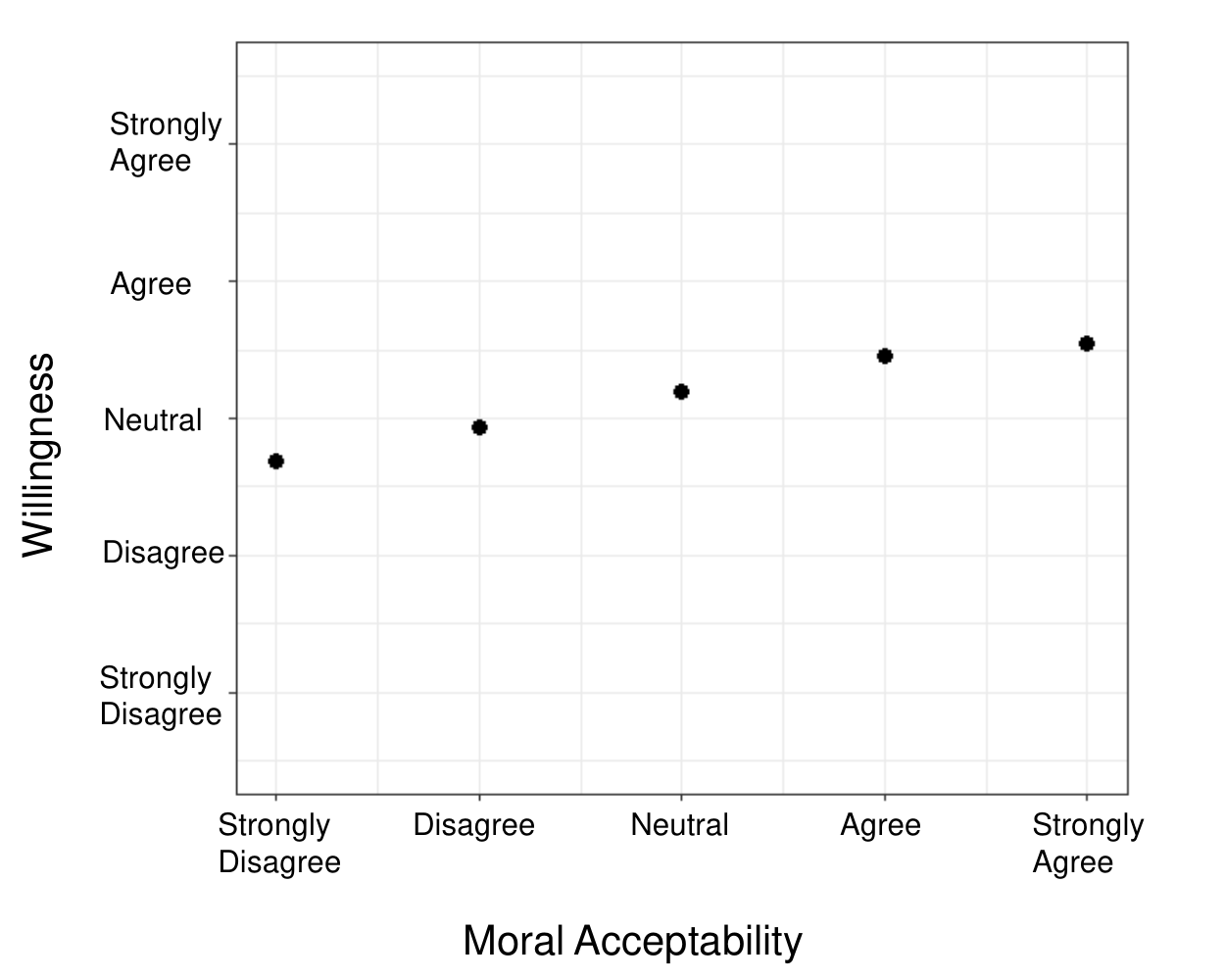}}
	\caption{\textbf{Moral acceptability is related to willingness to use Ceremode.} We observed a positive association between the moral acceptability of specific uses of Ceremode and willingness to use Ceremode for each purpose. Circles represent means at each value of moral acceptability, standard errors of the mean too small to observe at this scale.}\label{fig:moral}
	\centering
\end{figure}

Across all items, moral acceptability was positively related to the willingness to use Ceremode (See Table ~\ref{table:moral} and Fig.~\ref{fig:moral}). Post hoc t-tests at the item level indicate that subjects viewed repair to be more morally acceptable than enhancement overall (t(2771) = -13.14, $p< 1.0 \times 10^{-16}$), affecting core functions to be more morally acceptable than peripheral functions (t(2771) = -28.91, $p< 1.0 \times 10^{-16}$), and affecting themselves to be more morally acceptable than others (t(2771) = 3.762, $p < 0.001$). Importantly, the median response for each condition was to "Agree" that uses were morally acceptable, suggesting that differences between condition involve relative acceptability instead of acceptable \emph{versus} unacceptable uses.

\begin{table}[ht]
	\caption{\textbf{Moral acceptability associations with willingness to use Ceremode}} \label{table:moral}
	
	\begin{tabular}{l c c c c c} \toprule
		
		{} & {Estimate} & {Std. Error} & {df} & {t-value} & {\emph{p}} \\ \midrule
		
		Intercept  & 3.317 & 0.128 & 1037 & 25.838 &  $<$0.001\\
		Age  	   & -0.120 & 0.002 & 889 & -5.812 &  $<$0.001\\
		Education  & -0.048 & 0.034 & 890 & -1.418 &  0.156\\
		Gender     & 0.001 & 0.047 & 890 & 0.021 &  0.983\\
		Moral Acceptability & 0.160 & 0.009 & 2737 & 18.297 &  $<$0.001\\
		\bottomrule
		
	\end{tabular}
\end{table}

\clearpage
\newpage

There was no significant relationship between political affiliation and the willingness of individuals to enhance or repair cognitive functions (See Table~\ref{table:political}). Additionally, there was no significant relationship between neoliberalism and willingness to use Ceremode (See Table~\ref{table:neoliberal}). In addition, there was no significant interaction between openness to experience and core or
peripheral features (See Table~\ref{table:interactionsopenness})

\begin{table}[ht]
	\caption{\textbf{Main effects of political affiliation (Democrats as reference group)}} \label{table:political}
	
	\begin{tabular}{l c c c c c} \toprule
		
		{} & {Estimate} & {Std. Error} & {df} & {t-value} & {\emph{p}} \\ \midrule
		
		Intercept   & 3.997 & 0.129 & 887 & 30.891 &  $<$0.001\\
		Age  	    & -0.012 & 0.002 & 887 & -5.515 &  $<$0.001\\
		Education   & -0.040 & 0.034 & 887 & -1.165 &  0.244\\
		Gender      & 0.024 & 0.048 & 887 & 0.497 &  0.619\\
		Republican  & -0.034 & 0.062 & 887 & -0.537 &  0.591\\
		Independent & -0.062 & 0.058 & 887 & -1.070 &  0.285\\
		Other	    & 0.054 & 0.135 & 887 & 0.400 &  0.690\\
		\bottomrule
		
	\end{tabular}
\end{table}

\begin{table}[ht]
	\caption{\textbf{Main effect of the neoliberalism summary score}} \label{table:neoliberal}
	
	\begin{tabular}{l c c c c c} \toprule
		
		{} & {Estimate} & {Std. Error} & {df} & {t-value} & {\emph{p}} \\ \midrule
		
		Intercept      & 3.842 & 0.152 & 889 & 25.215 &  $<$0.001\\
		Age  	       & -0.012 & 0.002 & 889 & -5.620 &  $<$0.001\\
		Education      & -0.039 & 0.034 & 889 & -1.125 &  0.261\\
		Gender         & 0.016 & 0.048 & 889 & 0.335 &  0.738\\
		Neoliberalism  & 0.044 & 0.028 & 889 & -0.598 &  0.110\\
		\bottomrule
		
	\end{tabular}
\end{table}

\clearpage
\newpage
\begin{table}[ht]
	\caption{\textbf{Interaction model between openness to experience and trait fundamentalness}} \label{table:interactionsopenness}
	\begin{tabular}{l c c c c c} \toprule
		
		{} & {Estimate} & {Std. Error} & {df} & {t-value} & {\emph{p}} \\ \midrule
		
		Intercept   	 & 3.849 & 0.156 & 942 & 24.751 &  $<$0.001\\
		Age  	    	 & -0.119 & 0.002 & 889 & -5.622 &  $<$0.001\\
		Education  		 & -0.040 & 0.034 & 889 & -1.174 &  0.241\\
		Gender           & 0.021 & 0.048 & 889 & 0.432 &  0.666\\
		Core/Peripheral  & 0.087 & 0.053 & 2771 & 1.641 &  0.101\\
		Openness    	 & 0.032 & 0.026 & 1027 & 1.259 &  0.208\\
		Other	    	 & -0.021 & 0.013 & 2771 & -1.574 &  0.116\\
		\bottomrule
		
	\end{tabular}
\end{table}

\clearpage
\newpage

\section*{Discussion}
In self-report surveys measuring moral assessments of applications of brain stimulation on various cognitive functions, we found partial support for our hypotheses. Subjects were more willing to repair than enhance cognitive functions overall. Moreover, the reported moral acceptability of specific uses of Ceremode was associated with willingness to use Ceremode for that purpose. However, several results differed from our predictions. Contrary to our hypothesis, subjects were more willing to use Ceremode on others than themselves. In addition, individuals were most willing to repair rather than enhance core functions in others overall. Moreover, they expressed similar willingness to use Ceremode to influence peripheral features in themselves and others, but demonstrated willingness to influence the core features in others but \emph{not} on themselves.

Our findings suggest that moral evaluation contributes to willingness to use brain stimulation in zero-sum scenarios. When given vignettes indicating that a hypothetical brain stimulation technology comes in a limited supply, subjects prioritized repair over enhancement in alignment with their moral judgments. However, overall willingness to use brain stimulation on core or peripheral functions and on themselves rather than others was in the opposite direction of ratings of moral acceptability. This suggests that moral evaluation may contribute to judgments to repair over enhancing functions, but that that they override moral dispositions for the other conditions. In the case of self \emph{versus} other brain stimulation uses, significant main effects were observed in opposite directions: subjects may believe it is somewhat more appropriate to modify themselves, but report willingness to apply brain stimulation to others. Subject's high willingness to repair core functions in others may suggest that additional social dispositions override certain moral intuitions. Research shows that moral commitments can dissociate from moral judgments \cite{lombrozo2009role}, and our current study suggests that it will be important to  clarifying links between core moral commitments, discrete moral judgments, hypothetical willingness, and real-world behavior in brain stimulation uses. Because similar dissociations at the person-level and policy level have been observed in attitudes toward pharmaceuticals for enhancement, it is clear that systematic research should continue to assess specific moral attitudes and their relationships to hypothetical willingness and actual behavior. 

The most surprising finding -- that individuals were more willing to influence others, and specifically for core functions -- bears special consideration. In general, we know little about how the public views the potential resource distribution for cognitive enhancement. It is possible that individuals in the current study were altruistically inclined to repair others' ``true" selves when the self has been lost due to brain injury. Curiously, individuals were less inclined to endorse repair of core functions for themselves. Speculatively, this could be due to the difficulty in imagining what it is like to lose one's own authentic, holistic sense of self (the ``Mind's I" \cite{hofstadter2006mind}), and the relatively easier task of being able to observe how others change due to injuries in the public imagination or by experiences with friends and family. This potential link between imaginability and willingness could be explicitly tested in future studies. An additional nuance is worth noting: we did find some partial support for our prediction that individuals would favor self-enhancement for performance gains over others. Specifically, individuals were more likely to enhance and repair peripheral features in themselves relative to others. Since some of these ``peripheral" features include those that are valuable in a performance-based economy (e.g., memory, concentration), it is possible that individuals may prioritize these uses for personal gains over others. Examining individuals' beliefs about the links between specific functions and real-world outcomes could clarify this finding. Further, examining the relationships between moral commitments and moral judgment \cite{lombrozo2009role} could help explain why some individuals favor specific uses in themselves or others.

Interestingly, contrary to prior work in pharmaceuticals \cite{riis2008preferences, cabrera2015empirical}, we did not find that individuals were less willing to influence traits more ``core" to authentic identity with hypothetical brain stimulation. This indicates that consumer intuitions may differ due to the hypothetical mechanism of delivery. brain stimulation and pharmaceuticals can both be ``neuromodulatory" and influence cognition. However, brain stimulation is generally administered exogenously using electromagnetism, whereas pharmaceuticals are administered intra-orally and chemically modulate metabolic, neural, and other cerebrovascular effects. Moreover, brain stimulation predominantly emerged in the 1980s, whereas pharmaceutical manufacturing began in the late 1890s. This means that public awareness, acceptance, and reasoning about pharmaceuticals has developed for many decades longer than for brain stimulation (though see \cite{wexler2017recurrent} concerning an exception through the 1920s). How sociocultural processes shape evaluation of the specific mechanisms of interventions and their public or professional acceptance could inform an important new area of study for research pertinent to reasoning and public policy.

In this vein, our research contributes an important empirical basis from which research informing public education, outreach, and policy could develop. Several groups have opined that growing trends toward in brain stimulation could lead to issues with ethical and equitable distribution of clinical and enhancing or ``neuroaesthetic" \cite{chatterjee2011neuroaesthetics} uses of brain stimulation \cite{hamilton2011rethinking,medaglia2017brain}. Moreover, the public is more worried than enthusiastic about enhancement using invasive forms of brain stimulation to enhance themselves \cite{funk2016public}. There have also been efforts to educate and caution at home brain stimulation users about unknown or harmful effects \cite{wurzman2016open}. Thus, there are several stakes amenable to empirical inquiry. The first concerns the effects of specific uses of brain stimulation, where investigators must determine what functions are affected, to what degree, and in what circumstances \cite{pugh2017need}. The second, depending on the first, concerns whether and how ethical and equitable distribution are feasible \cite{hamilton2011rethinking}. The third, motivated by the current work, concerns how objective facts about brain stimulation and the ethical and equitable distribution of brain stimulation are viewed by the public and professional brain stimulation practitioners. This latter issue has heretofore largely been subject to intensive conceptual discussions and opinions, despite the fact that the public is demonstrably sensitive to many of the key debates and stakes\cite{cabrera2015reasons}. More pressingly, the objective study of public and professional attitudes and behaviors toward brain stimulation is required to make informed decisions about how these phenomena effect the actual practices of brain stimulation, and how we might seek to influence these processes \cite{wexler2017crowdsourced,wexler2017social,wexler2016practices,wexler2017who}. 

Concerning potential interventions targeting public or professional attitudes toward brain stimulation, our findings suggest at least two directions worth immediate inquiry. First, as described above, studying how judgments are influences by an intervention's mechanism of action and individuals' understanding of them and their objective effects is important. Second, we should seek to connect judgments about interventions such as brain stimulation to broader research in human decision making \cite{kahneman2011thinking}, neuroeconomics \cite{glimcher2013neuroeconomics}, and moral reasoning \cite{greene2002and}. Each of these fields are successfully clarifying how specific neurocognitive traits, biases, and adaptive processes interact to produce more or less rational and emotionally valenced responses and shape beliefs. While we focus on survey responses to zero-sum vignettes, we can also begin to connect judgments about self enhancement and repair to cognitive models, identify dimensions that can influence specific judgments, and explore analogous judgments of willingness in nonzero sum scenarios. This would let us test how moral acceptability may or may not be influenced to change judgments about brain stimulation, and whether other cognitive-emotional processes contribute and can be modified. Indeed, we found that individuals viewed affecting themselves with brain stimulation to be more morally acceptable, but were overall more willing to use brain stimulation on others to repair core functions essential to empathic, altruistic functioning.

\paragraph{Limitations}
Several limitations to the current study can and should be addressed with future research. The current study is subject to the limitations inherent in self-report survey research. We cannot identify causal influences with the current study. In addition, the structure of the survey did not allow us to test for possible order dependencies between reports of willingness and moral acceptability since the latter always followed the former. Future studies could manipulate moral cuing at the survey section or item level to test whether moral primacy influences willingness to use brain stimulation for specific purposes. In addition, it is well established that reverse coding and item-level issues can influence ratings \cite{weems2001impact}. Moreover, while we uncovered several significant main effects and interactions, there was notable variation in responses that was not accounted for by the variables we examined. An interesting future direction would be to examine item-level effects both in terms of accurately assessing attitudes and identifying targets for influencing judgments. In addition, while we included a specific mechanism of injury in our repair condition, other mechanisms of injury could be investigated as a dimension of interest. Another salient variable is the religious identity of individuals; in a public survey regarding invasive technologies, religious individuals are more reluctant than atheists and agnostics to be willing to receive enhancements \cite{funk2016public,wexler2017who}. Religious affiliation may interact with the variables explored here, including underlying moral evaluations. Finally, while prior research suggests that survey data collected via Mechanical Turk can be representative of data acquired in person \cite{buhrmester2011amazon,rand2012promise}, it is unknown whether this generalizes to the current context.

\section*{Conclusion}
Uses of brain stimulation are increasingly available, but empirical studies regarding about public attitudes and judgments have lagged behind opinions and policy. The current study can begin an effort to understand judgments and moral dispositions toward cognitive enhancement and repair. Our results suggest that while brain stimulation uses tend to be morally acceptable, the degree of acceptability informs its use to repair functions core to identity in others, but is at odds with the willingness to endorse specific uses. Future efforts can begin to apply these findings to contexts beyond brain stimulation to inform human judgment and reasoning more generally. We can begin to develop a mature understanding of moral judgments regarding the use of technologies to repair and enhance various cognitive functions with the aim of informing public and professional policy, education, and biomedical interventions.
	
\newpage
\subsection*{Acknowledgements}
The authors thank Anna Wexler for helpful comments on an early draft of the manuscript. JDM acknowledges support from the Office of the Director at the National Institutes of Health and the National Institute of Mental Health through grant number 1-DP5-OD-021352-01, R01-DC014960-01, and the Perelman School of Medicine. The content is solely the responsibility of the authors and does not necessarily represent the official views of any of the funding agencies. The funders had no role in study design, data collection and analysis, decision to publish, or preparation of the manuscript.

\clearpage
\newpage

\bibliographystyle{naturemag} 
\bibliography{../../../../LaTeX_Bibliography/JDMReferences}

\end{document}